\documentclass[letterpaper, preprint, paper,11pt]{AAS}	

\usepackage[utf8]{inputenc}

\usepackage{graphicx}
\usepackage{amsmath}
\usepackage[version=4]{mhchem}
\usepackage{siunitx}
\usepackage{longtable,tabularx}
\usepackage{float}
\usepackage{placeins}

\usepackage[utf8]{inputenc}
\usepackage{amssymb}
\usepackage{placeins}
\usepackage{bm}
\usepackage{siunitx}
\usepackage{graphicx}
\usepackage{subcaption}
\usepackage{placeins}
\usepackage{graphicx}
\usepackage{float}
\usepackage{footnpag}		
\usepackage{bm}
\usepackage{amsmath}
\usepackage{caption}
\usepackage{algorithm}
\usepackage{algpseudocode}
\usepackage{footnpag}			      	
\usepackage{comment}
\usepackage{xcolor}

\usepackage{tikz}

\usepackage{subcaption}

\PaperNumber{26-047} 

\title{Consensus Based Task Allocation for Angles-Only Local Catalog Maintenance of Satellite Systems\thanks{Approved for public release; distribution is unlimited. Public Affairs approval $\#$AFRL-2025-4455. This work is supported by the SMART SEED Innovation Award [Grant \#FA9453-21-2-0064].}}

\author{Harrison Perone,\thanks{Undergraduate Mechanical Engineering Student, University of Connecticut, Storrs, CT 06269}
\ Christopher W. Hays\thanks{Research Aerospace Engineer, Air Force Research Laboratory Space Vehicles Directorate, Kirtland AFB, NM. 87117} 
}

\begin{document}

\newcommand{\alex}[1]{\textcolor{blue}{#1}}
\newcommand{\kristina}[1]{\textcolor{teal}{#1}}
\newcommand{\ch}[1]{\textcolor{red}{#1}}

\newcommand{\reals}{{\mathbb R}}
\newcommand{\sphere}{{\mathbb S}}
\newcommand{\SOThree}{\textrm{SO}(3)}
\newcommand{\soThree}{\mathfrak{so}(3)}
\newcommand{\workspace}{W}
\newcommand{\nnreals}{\mathbb{R}_{\geq 0}}
\newcommand{\stateSpace}{\mathcal{X}}
\newcommand{\inputSpace}{\mathcal{U}}

\newcommand{\deputyState}{{\mathbf x}}
\newcommand{\deputyStateSpace}{{\mathcal X}}
\newcommand{\deputyStateMat}{{\mathbf A}}
\newcommand{\meanmotion}{\eta}
\newcommand{\Azimuth}{\text{Az}}
\newcommand{\Elevation}{\text{El}}

\newcommand{\todo}[1]{\par\noindent{\color{purple}\raggedright\textsc{#1}\par\marginpar{\large$\star$}}}

\newcommand{\chiefState}{{\mathbf R}}
\newcommand{\chiefStatex}{\mathbf{x}_0}
\newcommand{\chiefVels}{{\boldsymbol \omega}}
\newcommand{\chiefVel}{\omega}
\newcommand{\chiefInertia}{{\mathcal J}}
\newcommand{\chiefInertiaInv}{[{\mathcal J}]^{-1}}
\newcommand{\chiefInput}{u}
\newcommand{\traj}{\xi}
\newcommand{\dynfunc}{f}
\newcommand{\constraintfunc}{h}
\newcommand{\controlfunc}{c}
\newcommand{\initial}{\theta}

\newcommand{\inspectState}{\mathbf{y}}

\newcommand{\deputyUncertainty}{\mathbf{P}}
\newcommand{\noiseInput}{\mathbf{G}}
\newcommand{\processNoise}{\mathbf{Q}}
\newcommand{\obsMat}{C}
\newcommand{\obsState}{\mathbf{y}}
\newcommand{\pointing}{\mathbf{p}}
\newcommand{\fov}{\mathcal{X}_{\textrm{FOV}}}
\newcommand{\aov}{\alpha}
\newcommand{\maxEnt}{\mathcal{E}}

\newcommand{\skewMat}[1]{\mathcal{S}({#1})}
\newcommand{\idMat}[1]{\mathbf{I}_{#1 \times #1}}
\newcommand{\zeroMat}[2]{\mathbf{0}_{#1 \times #2}}
\newcommand{\emptyMat}[2]{\mathbf{\emptyset}_{#1 \times #2}}
\newcommand{\trace}{\mathrm{trace}}

\newcommand{\rotmat}[2]{\mathcal{R}_{#1 \to #2}}
\newcommand{\dotrotmat}[2]{\dot{\mathcal{R}}_{#1\to#2}}
\newcommand{\hillsframe}{\mathcal{H}}
\newcommand{\bodyframe}{\mathcal{B}}
\newcommand{\inertialframe}{\mathcal{I}}
\newcommand{\restr}{\downarrow}


\newcommand{\B}{\mathcal{B}}
\newcommand{\argmax}{\mathop{\mathrm{argmax}}\limits}
\newcommand{\cost}{\mathcal{J}}

\newcommand{\switchtime}{\tau}
\newcommand{\sunvector}{\mathbf{s}}

\newcommand{\eulang}{\Gamma}
\newcommand{\roll}{\phi}
\newcommand{\pitch}{\theta}
\newcommand{\yaw}{\psi}
\newcommand{\angularvel}[3]{\boldsymbol{\omega}_{#1#2}^{#3}}
\newcommand{\angularaccel}[3]{\boldsymbol{\alpha}_{#1#2}^{#3}}
\newcommand{\dotangularvel}[3]{\dot{\boldsymbol{\omega}}_{#1#2}^{#3}}
\newcommand{\torque}{\boldsymbol{\tau}}

\newcommand{\norm}[1]{|#1|}
\maketitle

\thispagestyle{plain}
\pagestyle{plain}

\begin{abstract}

In order for close proximity satellites to safely perform their missions, the relative states of all satellites and pieces of debris must be well understood. This presents a problem for ground based tracking and orbit determination since it may not be practical to achieve the required accuracy. Using space-based sensors allows for more accurate relative state estimates, especially if multiple satellites are allowed to communicate. Of interest to this work is the case where several communicating satellites each need to maintain a local catalog of communicating and non-communicating objects using angles-only limited field of view (FOV) measurements. However, this introduces the problem of efficiently scheduling and coordinating observations among the agents. This paper presents a decentralized task allocation algorithm to address this problem and quantifies its performance in terms of fuel usage and overall catalog uncertainty via numerical simulation. It was found that the new method significantly outperforms the uncertainty-fuel Pareto frontier formed by current approaches.
\end{abstract}


\section{Introduction}


As the number of objects in LEO continues to grow, so too does the importance of maintaining accurate and up to date state information about neighboring objects. This information enables operators (or the satellites themselves) to make more informed decisions about collision avoidance maneuvers. Additionally, it allows for a wider variety of close proximity mission sets such as orbital construction, inspection, and formation flying \cite{space_construction, q_learning_tumbling_inspection, UKF_formation_flying}.

One option for addressing the need for more accurate relative state estimates is by using a collection of communicating agents equipped with limited FOV angles-only sensors. Each agent is tasked with maintaining its own local catalog of the states and covariances of objects within the local operating environment. The goal is that the uncertainty (some function of the covariance) of all objects is below a predefined threshold in each catalog. This approach presents several subproblems, namely control of each agent, distributed state estimation, construction of the communication graph, and observation task allocation. These problems and potential solutions are studied in Hays et al. \cite{hays2024}. The specific contribution of this paper is to improve upon the task allocation algorithm from that work in terms of fuel expenditure and the ability to maintain the uncertainty of all objects below a desired threshold. Additionally, the new algorithm does not require a centralized supervisor meaning that it is more robust and reduces communication requirements.

With the increasing importance of autonomous multi-agent systems like satellites and UAVs, there has been a corresponding interest in efficient and effective task allocation algorithms for these systems. For instance, Choi et al. \cite{choi2009} presents a general-purpose algorithm for decentralized task allocation of multi-agent systems called the Consensus-Based Bundle Algorithm (CBBA). Several follow-up works have looked at modifications to this algorithm that make it better suited for dynamic environments. Johnson et al. \cite{johnson_asynchronous_2011} presents an asynchronous version of the algorithm (called ACBBA) that is much more practical for real multi-agent systems. Additionally, Buckman et al. \cite{buckman_partial_2019} focused on allowing the agents to incorporate a new task into their plans without completely re-planning. There have also been several studies that demonstrate the effectiveness of this algorithm for multi-agent systems via simulation. For instance, variants of the CBBA approach were tested for UAVs swarms \cite{bertuccelli_UAV, zhang_UAV}. Its performance has also been assessed for Earth observation missions with satellites \cite{aguilar_jaramillo_decentralized_2025, phillips_case_2021, lee_agile_observation}. All of these studies used dynamic or complex tasks of some form. These include time-varying task locations and values, heterogeneous agents and tasks, limited time availability of tasks, and dependencies between tasks. To the best of the authors' knowledge, the feasibility of the CBBA algorithm for the catalog maintenance problem has not yet been assessed. However, given that the problem has some of the complexities looked at in these studies, the CBBA algorithm represents a promising research direction.

There are several factors that add to the complexity of designing a task allocation algorithm for the local catalog maintenance problem. First and foremost is that the states and covariances of all objects are time-varying. Any algorithm scheduling future tasks either needs to simulate the state of the entire catalog of objects forward in time or assume that changes occur slowly enough to be considered quasi-static for planning purposes. Additionally, it is not obvious when an observation task should be considered ``complete." The naive approach calls a task complete as soon as that object falls below the uncertainty threshold. However, because uncertainty monotonically increases with time, tasks that were just completed would quickly rise back above the threshold and need follow-up observations. Finally, because there is communication between the agents, when one agent observes an object, its uncertainty will decrease for all of the agents. This means the observation tasks assigned to each agent will affect the quality of observation tasks assigned to other agents. All of these issues are addressed with the algorithm developed in this work. Every time a target is observed, a modified version of CBBA plans a small number of observations into the future based on the current state of the system. This frequently revises the plans of the agents based on the current state of the system and the tasks chosen by other agents. Observation tasks are considered complete when the rate of decrease of a new observation score function falls below a certain value.


\section{Background}

\subsection{Notation and Definitions}
\label{sec:Notation and Definitions}

Let $\mathbb{R}$ be the set of real numbers and $\mathbb{N}$ be the set of natural numbers. Next, let $\mathbb{R}^n$ be the set of n dimensional real-valued column vectors, and $\mathbb{R}^{n\times m}$ be the set of real-valued matrices in with n rows and m columns. The matrix $\mathbf{I}_n\in\mathbf{R}^{n\times n}$ is the n-dimensional identity matrix. Given a vector $\mathbf{x}^\mathcal{H}$ expressed in frame $\mathcal{H}$, it can be converted to a different frame $\mathcal{B}$ with the following rotation matrix $\mathbf{x}^\mathcal{B}=\mathbf{R}_{\mathcal{H}}^\mathcal{B}\mathbf{x}^\mathcal{H}$. The angular velocity of frame $\mathcal{B}$ relative to frame $\mathcal{H}$ when expressed in frame $\mathcal{I}$ is given by $\boldsymbol{\omega}_{\mathcal{H}\mathcal{B}}^\mathcal{I}$. The cross product operations is defined as follows for a vector $\mathbf{v}=[v_1,v_2,v_3]^\top\in\mathbb{R}^3$

$$(\mathbf{v})^\times=
\begin{bmatrix}
0&-v_3&v_2\\
v_3&0&-v_1\\
-v_2&v_1&0
\end{bmatrix}
$$

Its name comes from the fact that the cross product of two vectors $\mathbf{v}$ and $\mathbf{w}$ can also be written as the matrix multiplication $(\mathbf{v})^\times \mathbf{w}$ where $\mathbf{w}\in\mathbf{R}^3$. The Euclidean norm of a vector $\mathbf{v}$ is given by $||\mathbf{v}||=\sqrt{\mathbf{v}^\top\mathbf{v}}$. For matrices $\mathbf{A}$ and $\mathbf{B}$, the Kronecker product is represented by the $\otimes$ symbol and performs the following operation:

$$A\otimes B=
\begin{bmatrix}
a_{11}\mathbf{B}&a_{12}\mathbf{B}&\ldots&a_{1n}\mathbf{B}\\
a_{21}\mathbf{B}&a_{22}\mathbf{B}&\ldots&a_{2n}\mathbf{B}\\
\vdots&\vdots&\ddots&\vdots\\
a_{m1}\mathbf{B}&a_{m2}\mathbf{B}&\ldots&a_{mn}\mathbf{B}\\
\end{bmatrix}$$

\subsection{Relative Orbital Dynamics}
\label{sec:Relative Orbital Dynamics}

The Clohessy–Wiltshire–Hill (CWH) equations \cite{CWH_Equations} are a set of linear differential equations that can be used to describe relative orbital dynamics. These equations apply to the non-inertial Hill frame in which a satellite considered the chief is in a circular orbit around the parent body and is used as the origin of the frame. The x axis points away from the center of the parent, the y axis points in the chief's direction of motion, and the z axis completes the right-handed triad. The relative state vector $\mathbf{x} = [\mathbf{r}_x,\mathbf{r}_y,\mathbf{r}_z,\mathbf{v}_x,\mathbf{v}_y,\mathbf{v}_z]^{\top}$ is used to describe the relative position $\mathbf{r}$ and relative velocity $\mathbf{v}$ of a nearby object in the Hill frame. This state vector evolves according to the following uncontrolled dynamics:

\begin{align}
\begin{split}
&\dot{\mathbf{v}}_x=2\eta\mathbf{v}_y+3\eta^2\mathbf{r}_x \\
&\dot{\mathbf{v}}_y=-2\eta\mathbf{v}_x \\
&\dot{\mathbf{v}}_z=-\eta^2\mathbf{r}_z
\end{split}
\end{align}

where $\eta$ is the constant angular velocity of the chief around the parent body. These dynamics can also be written in the form $\dot{\mathbf{x}}=\boldsymbol{\Phi}\mathbf{x}$ where $\boldsymbol{\Phi}$ is given by:

\begin{equation}
\boldsymbol{\Phi}=
\begin{bmatrix}
0&0&0&1&0&0\\
0&0&0&0&1&0\\
0&0&0&0&0&1\\
3\eta^2&0&0&0&2\eta&0\\
0&0&0&-2\eta&0&0\\
0&0&-\eta^2&0&0&0\\
\end{bmatrix}
\end{equation}

This allows the dynamics from Eq. (2) to be generalized to multi-satellite systems with $\dot{\mathbf{x}}=(\mathbf{I}_N\otimes\mathbf{\Phi})\mathbf{x}$ where $\mathbf{x}$ in this context is the combination of all satellite states $\mathbf{x}=[\mathbf{x}_1^{\top},...,\mathbf{x}_N^{\top}]^{\top}$.

\subsection{Natural Motion Trajectories}
\label{sec:Natural Motion Trajectories}

Natural Motion Trajectories (NMTs) are a periodic set of solutions to the CWH dynamics presented in Eq. (1). They are useful in a variety of close proximity mission sets because they allow for predictable relative motion without any fuel expenditure. There are several types of paths that NMTs can form in the Hill frame including points, lines, ellipses, and spirals. Each of these must satisfy the following initial condition $\mathbf{v}_y=-2\eta\mathbf{r}_x$. Additionally, each type has its own initial condition requirements. For a point, the only nonzero term is $\mathbf{r}_y$. This orbits ahead of or behind the chief in its direction of motion. In the case of a line, the non-zero initial conditions are $\mathbf{r}_z=c\sin(\psi)$ and $\mathbf{v_z}=\eta c\cos(\psi)$ where c is the amplitude of the line and $\psi$ is the initial angle relative to the xy plane. This NMT oscillates above and below the xy plane at a fixed y coordinate. Finally, an ellipse must satisfy $\mathbf{v}_x=(\eta/2)\mathbf{r_y}$. Note that the z components can be chosen freely meaning these ellipses can have arbitrary inclinations relative to the xy plane.

\subsection{Attitude Dynamics}
\label{sec:Attitude Dynamics}

This work assumes that all agents are able to apply control torques in order to view targets of interest. The attitude of the agent's body frame relative to the inertial frame can be described by the Euler angles $\Gamma=[\phi,\psi,\gamma]^{\top}$ where $\phi$, $\psi$, and $\gamma$ represent the yaw, pitch, and roll angles respectively. The ZYX rotation sequence is chosen so that yaw and pitch angles directly relate to the azimuth and elevation coordinates used for observing targets. In other words, $\mathbf{R}_{\mathcal{B}}^{\mathcal{I}}=\mathbf{R}_{\mathcal{B}}^{\mathcal{I}}(\gamma)\mathbf{R}_{\mathcal{B}}^{\mathcal{I}}(\psi)\mathbf{R}_{\mathcal{B}}^{\mathcal{I}}(\phi)$. The agent's angular velocities and control torques are given by $\boldsymbol{\omega}_{\mathcal{IB}}^\mathcal{B}$ and $\boldsymbol{\tau}^\mathcal{B}$ respectively. The rotational kinematics of the system are given by:

\begin{equation}
\dot{\mathbf{R}}_{\mathcal{B}}^{\mathcal{I}}=\mathbf{R}_{\mathcal{B}}^{\mathcal{I}}(\boldsymbol{\omega}_{\mathcal{IB}}^\mathcal{B})^{\times}
\end{equation}
\begin{equation}
\dot{\Gamma}=
\begin{bmatrix}
-\cos(\phi)\tan(\psi)&-\sin(\phi)\tan(\psi)&-1\\
\sin(\phi)&-\cos(\phi)&0\\
-\cos(\phi)\sec(\psi)&-\sin(\phi)\sec(\psi)&0
\end{bmatrix}
\boldsymbol{\omega}_{\mathcal{IB}}^\mathcal{B}
\end{equation}

The rotational dynamics of the system can be found using Euler's equation for rigid bodies.

\begin{equation}
\dot{\boldsymbol{\omega}}_{\mathcal{IB}}^\mathcal{B}=-\mathcal{J}^{-1}(\boldsymbol{\omega}_{\mathcal{IB}}^\mathcal{B})^\times[\mathcal{J}\boldsymbol{\omega}_{\mathcal{IB}}^\mathcal{B}]+\mathcal{J}^{-1}\boldsymbol{\tau}^\mathcal{B}
\end{equation}

Where $\mathcal{J}$ is the constant inertia matrix of the agent. The dynamics for the angular velocity of the body frame relative to the Hill frame can also be found using properties of angular velocities and the Hill frame itself. A full derivation is presented in \cite{hays2024}.

\begin{align}
\begin{split}
\dot{\boldsymbol{\omega}}_{\mathcal{HB}}^\mathcal{B}=-\mathcal{J}^{-1}((\boldsymbol{\omega}_{\mathcal{HB}}^\mathcal{B})^\times[\mathcal{J}\boldsymbol{\omega}_{\mathcal{HB}}^\mathcal{B}]+(\boldsymbol{\omega}_{\mathcal{HB}}^\mathcal{B})^\times[\mathcal{J}\boldsymbol{\omega}_{\mathcal{IH}}^\mathcal{B}]\\+(\boldsymbol{\omega}_{\mathcal{IH}}^\mathcal{B})^\times[\mathcal{J}\boldsymbol{\omega}_{\mathcal{HB}}^\mathcal{B}]+(\boldsymbol{\omega}_{\mathcal{IH}}^\mathcal{B})^\times[\mathcal{J}\boldsymbol{\omega}_{\mathcal{IH}}^\mathcal{B}])\\+
(\boldsymbol{\omega}_{\mathcal{HB}}^\mathcal{B})^\times\boldsymbol{\omega}_{\mathcal{IH}}^\mathcal{B}+
\mathcal{J}^{-1}\boldsymbol{\tau}^\mathcal{B}
\end{split}
\end{align}

\subsection{Principal Axes}
\label{sec:Principal Axes}

A covariance matrix is one way of expressing the spread of an $n$ dimensional gaussian distribution. An alternative description of the distribution is a set of orthogonal vectors corresponding to the principal axes of the hyper-ellipsoid. If the magnitudes of each of these vectors are assigned to the variance of the distribution along that axis, then these vectors form a complete description of the distribution around a given mean. Although they describe the same information, these vectors give a clearer understanding of how the uncertainty is distributed as opposed to the entries of the covariance matrix. 

To find this alternate description of a covariance matrix $\mathbf{P}\in\mathbb{R}^{n\times n}$, find the eigenvectors and eigenvalues of $\mathbf{P}$ such that: 
\begin{equation}
    \mathbf{P}\mathbf{u}_k=\lambda_k\mathbf{u}_k
\end{equation}
Where each $\mathbf{u}_k$ is the direction of the $k^{th}$ principal axis and each $\lambda_k$ is the corresponding variance along that axis.

\subsection{Consensus-Based Bundle Algorithm (CBBA)}
\label{sec:Evaluation}

The CBBA algorithm \cite{choi2009} is a decentralized multi-agent task allocation algorithm that can efficiently generate a near optimal solution. In addition to this, it still produces solutions when the agents have differing situational awareness. The problem that CBBA was designed to solve can be stated as follows. There are $N$ agents, $N_t$ tasks, and a maximum number of tasks per agent of $L_t$ (the planning depth). Every agent has an ordered list of tasks that it plans to complete known as its path. A path score function determines the quality of an agent's chosen path. The ultimate goal of the algorithm is to maximize the sum of all path scores across all agents without assigning multiple agents to the same task. CBBA approaches this problem by alternating between bundle construction and conflict resolution phases. The bundle construction phase involves each agent sequentially adding tasks to its path until no more tasks are available. Since bundle construction does not involve any communication, the paths created in this process often overlap with each other. The purpose of the conflict resolution phase is to resolve this by sending messages between pairs of agents. Each agent decides what to do with received information about a particular task by following a set of communication rules.


\section{Approach}\label{sec: Approach}


Hays et al. \cite{hays2024} use a fixed hysteresis time to approach the problem of deciding when a task is completed. It works by moving on to a new target if the current target object is below the uncertainty (Shannon Entropy) threshold in the catalog and it has been under observation by this agent for at least the length of the hysteresis time. If this is the case, then the agent chooses whatever object currently has the highest uncertainty to be its new observation target. This solution is easy to implement and can be effective for the local catalog maintenance problem. However, there are several ways that this algorithm can be made more observation and fuel efficient. For instance, considering the angular distance to observation targets, coordinating with neighboring agents to find a better global solution, and refining the function that scores the quality of a particular observation target. All of these will be addressed in the remainder of this section.


In order to meaningfully approach this problem, several constraints were placed on the design of this task allocation algorithm. For instance, it only uses the current state of the system when making tasking decisions. This significantly reduces the amount of computation required at the cost of choosing less optimal tasks, especially over long time horizons. Additionally, rather than looking for a continuous observation schedule, each agent will just find an ordered list of tasks that it currently plans on completing. This is because it is difficult to predict how long an observation should take without simulating how the entire system will evolve. Lastly, none of the agents can share tasks at a given time. This prevents all of the agents from just looking at the handful of objects that currently have the greatest uncertainty.

It is instructive to start by looking at the scoring function developed for this problem. It is used to determine when a target switch should occur and to decide what targets are most useful to which agents. Note that even though the goal of every agent is to keep the Shannon Entropy of the covariance matrix below a certain threshold value, this metric is not ideal for choosing observation targets since it is independent of the range to the target. The range is relevant because for an angles-only sensor with fixed angular uncertainty, the corresponding positional uncertainty will be proportional to the range to the target. Additionally, Shannon Entropy does not consider the way this uncertainty is distributed or the direction of observation. To see why this is relevant, consider the case where an elongated ellipsoid (corresponding to the position probability distribution) is aligned with the direction of observation. Since most of the uncertainty lies along the unobservable direction, the amount of information that can be gained from this observation is much lower than if the long axis of the distribution were perpendicular to the direction of observation. The following scoring metric aims to make use of this information for a given agent and target where $\mathbf{r}$ is the position vector of the target relative to the agent and $\bar{\mathbf{r}}=[\mathbf{r}^\top,\mathbf{r}^\top]^\top$.

\begin{equation}
\theta_k=\cos^{-1}\left(\frac{\bar{\mathbf{r}}\cdot \mathbf{u}_k}{||\bar{\mathbf{r}}||||\mathbf{u}_k||}\right)
\end{equation}

\begin{equation}
S=\sum_{k=1}^6\frac{\lambda_k\sin(\theta_k)}{||\mathbf{r}||}
\end{equation}

Where $\mathbf{u}_k$ and ${\lambda}_k$ are given by equation (7) for the covariance matrix of the target $P$. Eqs. (8) \& (9) aim to assign high scores to agent-target pairs that would lead to a large decrease in uncertainty when observed. They can be better understood by considering the three-dimensional case where $P$ is just the covariance of position and $\bar{\mathbf{r}}$ is simply $\mathbf{r}$. With this setup, each $\theta_k$ is the angle between the relative position vector and one of the three principal axis of the target's position distribution. When one of these angles is near zero or $\pi$, it means that it will be difficult to reduce the uncertainty of the target along this axis because it is close to the unobservable direction. Conversely, observations are most useful when $\theta_k\approx\pm\frac{\pi}{2}$. For the full six by six case presented in (8) and (9), the only change is that velocity becomes part of the covariance matrix and $\mathbf{r}$ becomes $\bar{\mathbf{r}}$ to match the dimensions. This still makes logical sense because it is also difficult to estimate component of velocity that is in the relative direction of the target. 

The total score S presented in Eq. (9) is the summation of the scores for each principal axis of $P$. Each term is scaled by $\sin(\theta_k)$ to assign higher scores to axes that do not align with the unobservable direction. Additionally, $\lambda_k$ scales each term by the variance along that axis. This means that axes (and hence targets) that are more uncertain receive higher scores. Lastly, the $\frac{1}{||\mathbf{r}||}$ term assigns higher scores to targets that are at closer range. As mentioned earlier, the positional measurement uncertainty for these targets is lower. Each of these factors values targets that an agent can gain a significant amount of information from, either because the target itself is uncertain, or because the agent is well positioned to take a measurement. Next, the algorithm that decides when it is time for an agent to switch targets is presented. Note that every agent runs this algorithm at every timestep.

\begin{algorithm}[t]
\caption{Target switching logic for the $i^{th}$ agent}
\begin{algorithmic}[1]

\State targetBlacklist = []
\State agentBlacklisted = False
\For{$j = 1$ to $N$} 
    \State $\hat{\mathbf{r}} = \hat{\mathbf{r}}_j - \hat{\mathbf{r}}_i$
    \State $\hat{\bar{\mathbf{r}}}=[\hat{\mathbf{r}}^\top,\hat{\mathbf{r}}^\top]^\top$
    \State $S_j=0$
    \For{$k = 1$ to 6}
    \State$\theta_k=\cos^{-1}\left(\frac{\hat{\bar{\mathbf{r}}}\cdot \mathbf{u}_k}{||\hat{\bar{\mathbf{r}}}||||\mathbf{u}_k||}\right)$
    \State$S_j=S_j+\frac{\lambda_k\sin(\theta_k)}{||\hat{\mathbf{r}}||}$
    \EndFor

    \State $E_{j}=\frac{n}{2}(1+\log{(2\pi)})+\log{|\mathbf{P_j}|}$
    \State $R_{j}=|\hat{\mathbf{r}}|$
    \State $D_{j} = (S_{j} - S_{j,\text{previous}})/\Delta t$ \Comment{Derivative of target score}
    \If {$E_{j} < \epsilon$}
        \State append j to $\text{targetBlacklist}$
    \EndIf
    \EndFor
    \If {agent i has an observation target}
        \State Let i's observation target be t
        \If {t is not in i's FOV}
            \State agentBlacklisted = True
        \ElsIf {$\neg (E_{j} < \epsilon, \forall j$) $\wedge$ ($D_{t}$ $> -\frac{\alpha}{R_{t}})$}
            \State targetSwitch = True
            \State append t to $\text{targetBlacklist}$
        \Else
            \State \text{targetSwitch = False}
        \EndIf        
    \EndIf

\end{algorithmic}
\end{algorithm}

$S_j$ is the score of the $j^{th}$ object from the perspective of agent $i$. The Shannon Entropy of $P_j$ is notated as $E_j$ and $n$ is the dimension of $P_j$. This algorithm makes use of a ``blacklist" that aims to limit unnecessary target switches. The targetBlacklist is a list of targets that an agent can not add to its path. Targets are added to agent $i$'s targetBlacklist when they fall below the entropy threshold $\epsilon$. This is because the states of these targets are considered sufficiently well understood. Additionally, entire agents can be considered ``blacklisted". This is implemented with the agentBlacklisted boolean that, if true, prevents agent $i$ from changing its current target. It is set to true when agent $i$'s target is not within the FOV of its sensor. This prevents agents from switching targets while rotating to their current target. Not only do they miss out on the information they could have gained from their original target, but they also expend a significant amount of fuel to change direction. Note that both of these concepts were implemented by modifying the CBBA algorithm to set certain path scores to zero based on the values of the targetBlacklist and agentBlacklisted. Lastly, the CBBA path scores are found by multiplying the observation score (from Eqs. (8) and (9)) of each task in an agent's path list by an exponentially decaying function of the total angular distance traveled to reach this task and summing over the items in the list.
\begin{equation}
\tilde{S}=\sum_{i=1}^mS_ie^{-L_i\mu }
\end{equation}
Where $m$ is the current number of tasks in the plan, $L_i$ is the total angular distance traveled to get from task one to task $i$, and $\mu$ is the discount factor.
The result is that this path scoring function assigns high scores to paths that observe high value targets over short paths.

When targetSwitch is True for agent $i$, it shares this information throughout the communication graph so that all agents can clear their bundle and path lists. Note that the current target is also added to agent $i$ targetBlacklist to prevent it from being reselected. Now CBBA can be run to optimize task allocation for the new state of the system. Finally, each agent sets their observation target to the first item in their new path list (it keeps its current target if it does not have any items). It is important to recognize that coordination across a multi-agent network is a non-trivial task, especially if the topology of the communication graph is time-varying. That said, this work does not present a new decentralized communication protocol since it should not significantly impact task allocation performance.

For attitude control of each of the agents Hays et al. used model predictive control (MPC) \cite{hays2024}. One of the main advantages of MPC is that it allows for state and control constraints to explicitly specified. However, this comes at the cost of frequent re-optimization which can be computationally expensive. Since the main goal of this work is to numerically evaluate the performance of various task allocation algorithms, it was decided that a proportional controller with a damping term would be better suited for this use case. For a given agent and target, the control law is given by the following:

\begin{equation}
\boldsymbol{\tau}=k_p\alpha\mathbf{t} + k_d(\boldsymbol{\omega}_t-\boldsymbol{\omega})
\end{equation}
\begin{equation}
\alpha = \cos^{-1}\left(\frac{\mathbf{r}_s\cdot\mathbf{r}}{||\mathbf{r}||}\right)
\end{equation}
\begin{equation}
\mathbf{t}=\mathbf{r}_s\times\frac{\mathbf{r}}{||{\mathbf{r}}||}
\end{equation}
\begin{equation}
\boldsymbol{\omega}_t=\frac{\mathbf{r}\times\mathbf{v}}{||\mathbf{r}||^2}
\end{equation}

Where $\mathbf{r}_s$ is the unit vector corresponding to the pointing direction of the sensor and $\boldsymbol{\omega}$ is the angular velocity of the agent. Eq. (14) finds the apparent angular velocity of the target relative to the agent as it moves across the sky. With this setup, the proportional term in equation (10) always applies a torque that turns the sensor towards the target. The damping term tries to match the angular velocity of the target so that targets with constant angular velocity can be tracked with zero steady state error.

Lastly, note that state estimation and communication graph construction are handled by the algorithms studied in \cite{hays2024}. Specifically, a Networked Distributed Kalman Estimator is used for state estimation. This allows agents to update their state estimates with their own measurements and measurements sent to them by other agents. Additionally, the state estimates themselves can be shared and then fused using Inverse Covariance Intersection (ICI) \cite{NOACK201735}. The communication graph describes which agents are able to share state or measurement information at a given time. It is chosen such that the agents can maintain state omniscience. In other words, their current communication links must be sufficient for the error of the state estimates of all objects to approach zero as time goes to infinity.

\section{Results and Discussion}
\label{sec:results}

\subsection{Evaluation}
\label{sec:Evaluation}

There are two important performance metrics for task allocation algorithms attempting to solve this problem, namely, fuel expenditure and the ability to keep the uncertainties of all objects below the threshold value. Total fuel expenditure can be quantified by taking the integral of each agent's control command signal and summing over all agents. In other words, total fuel expenditure is given by:
\begin{equation}
F=\sum_{i=1}^{N}\int_0^{T}{||\bm{\tau}_i||}\, dt
\end{equation}
Where $T$ simulation end time and $\bm{\tau}_i$ is the control vector for the $i^{th}$ agent. How well the algorithm keeps objects below the threshold is slightly more difficult to quantify. This is because checking if one or all of the agents are below the uncertainty threshold at a given time is just a binary metric. A more comprehensive way to evaluate this is by integrating the portion of each uncertainty signal that is above the threshold and summing the results.

\begin{equation}
h_{i,j} =
\begin{cases}
    0, & \text{if } E_{i,j}-\epsilon < 0 \\
    E_{i,j}-\epsilon,  & \text{otherwise}
\end{cases}
\end{equation}
\begin{equation}
C=\sum_{i=1}^{N}\sum_{j=1}^{O} \int_0^{T}{h_{i,j}}\,dt
\end{equation}

Where $E_{i,j}$ is the Shannon Entropy of object $j$ in agent $i$'s catalog, O is the number of objects, and N is the number of agents. An agent that performs well on this metric would be one that quickly reduces the uncertainty of all objects in its catalog and rarely lets them go back above the threshold. A task allocation algorithm that addresses the central problem of this paper should ideally minimize both fuel expenditure and clipped integral metrics.

\subsection{Simulation Results}
\label{sec:Results}

A simulation environment was set up in Python to test the effectiveness of the task allocation algorithm presented in this paper. For the sake of simplicity, this simulation assumes that all other agents are instantly notified when a task is completed rather than sending this information over the graph at each time step. The true communication process takes place over relatively short time scales meaning the impact on the final task allocation performance should be negligible. The simulation has two communicating agents that are both in elliptical NMT around the virtual chief. The initial coordinates of each agent were randomly set between -100 m and 100 m in each axis. It also had eight bodies whose initial positions were chosen with the same procedure, but whose velocities do not put them on closed natural motion trajectories. To ensure the agents were able to find their intended targets, the state estimates were initialized to the true states. The covariance matrix was set by giving all initial positions a variance of $10^1$ m and all velocities a variance of $10^{-3}$ m. The Kalman filter was given a process noise matrix of $10^{-3}\;\mathbf{I}_6$ and a measurement noise matrix of $10^{-4}\;\mathbf{I}_3$. All agents are equipped with one angles-only sensor that has a FOV of $10^{\circ}$. All simulations were run for 200 seconds with a step size of one second. It is worth noting that each of the algorithms was tested using the same set of random seeds. This means that the differences between them should be due to the different algorithms rather than randomness introduced by the initial conditions. In order to gauge the performance of the task allocation algorithm presented in this paper, its performance on both metrics is assessed over a range of algorithm parameters. Then, the best performing result will be compared to the hysteresis algorithm over a range of hysteresis parameters. Firstly, the effect of CBBA planning depth (the maximum number of tasks in an agent's plan) will be evaluated.

\begin{figure}[H]
    \centering
    \includegraphics[width=0.75\textwidth]{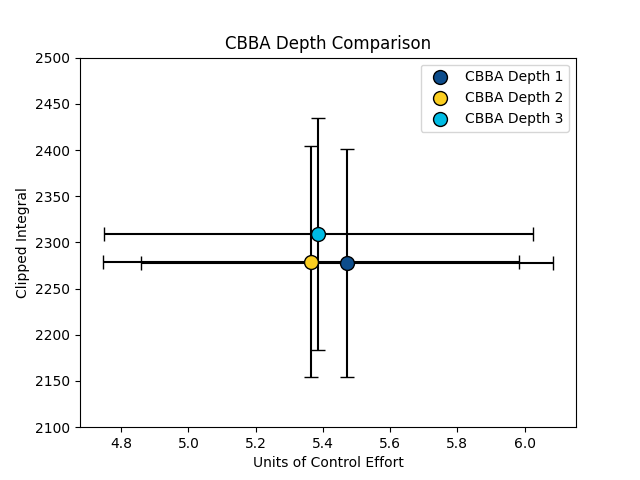}
    \caption{Scatter plot of mean performance for CBBA algorithm with different planning depths over a range of random initial conditions. For each configuration, 100 simulations were run, and the 95\% confidence interval is shown for each point with error bars.}
\end{figure}

Figure 1 shows that as CBBA planning depth increases, there is no statistically significant impact on the mean performance of the algorithm for either metric. This contradicted initial expectations since greater CBBA depth should create plans that consider the effects of completing a longer sequence of tasks. Therefore, at least in theory, increasing the depth should allow for more efficient routes between targets and use less fuel. However, there are several competing factors that better explain the results from Figure 1. For instance, once all of the targets have been brought below the threshold, Algorithm 1 will only allow a switch to occur if a target rises above the threshold again. This means that only one target will be off of the blacklist and available for observation. This would make planning depth irrelevant in this case. Now consider the opposite case where most of the targets are above the threshold (at the start of the simulation). The fact that the agents can plan to the full depth should improve their performance. However, the clipped integral metric is very sensitive to how long it takes for a target to be observed for the first time since it will be high above the threshold until then. Therefore, the optimal strategy during this phase likely puts lots of emphasis on observing high priority targets quickly rather than finding good routes between them. Lastly, frequent re-planning may mean that plans beyond a certain depth are unlikely to occur. This limits the benefit of planning to that depth in the first place if most go unused. Next, the effect of the discount factor for the CBBA algorithm will be evaluated.

\begin{figure}[h]
    \centering
    \includegraphics[width=0.75\textwidth]{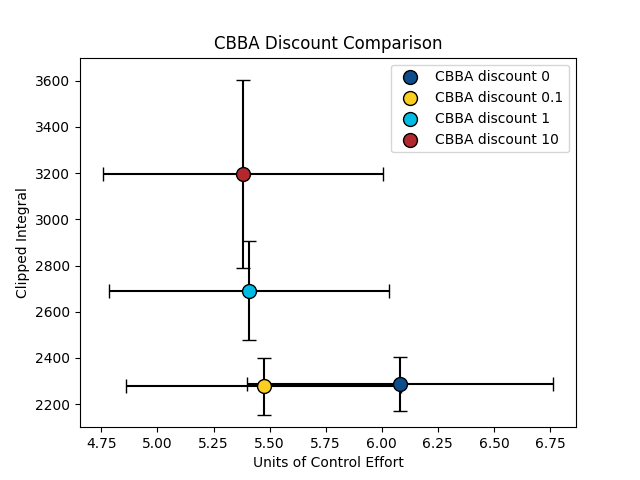}
    \caption{Scatter plot of mean performance for several CBBA discount values over a range of random initial conditions. For each configuration, 100 simulations were run, and the 95\% confidence interval is shown for each point with error bars.}
\end{figure}

\begin{figure}[H]
    \centering
    \includegraphics[width=0.75\textwidth]{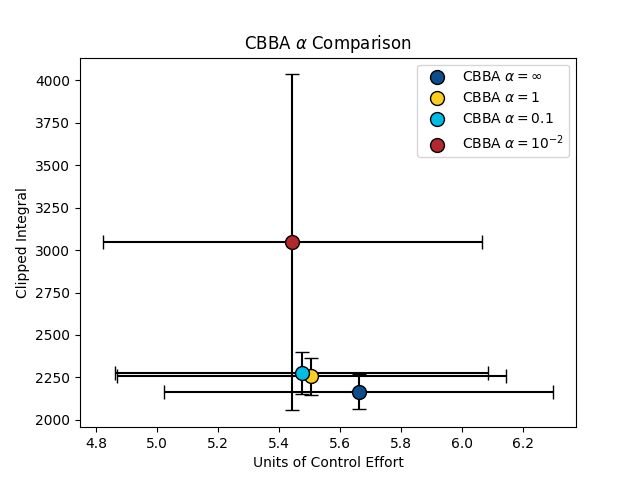}
    \caption{Scatter plot of mean performance for several CBBA $\alpha$ values over a range of random initial conditions. For each configuration, 100 simulations were run, and the 95\% confidence interval is shown for each point with error bars.}
\end{figure}

Figure 2 shows that there is a very abrupt change when the discount factor is roughly 0.1. At this point, increasing it leads to significant increases in clipped integral without much decrease in fuel use. Conversely, decreasing it increases control effort without any decrease in clipped integral. Therefore, a discount factor of 0.1 was chosen for this setup. Intuitively, this figure demonstrates the tradeoff between target value and angular distance. When the discount is zero, all targets have the same value regardless of the distance leading to high fuel cost but low clipped integral values. This is as opposed to high discount factors which almost exclusively look at the closest targets which leads to high clipped integral values but low fuel usage. The last parameter of interest has to do with the switching function itself. This is the $\alpha$ value and controls the rate of decrease that a switch is allowed to occur at.

Figure 3 is similar to Figure 2 in the sense that it forms the same rough ``L" shape and changes direction when $\alpha\approx0.1$. However, in this case the trend is reversed with $\alpha$ values increasing while moving down and to the right. This can be explained by the fact that large $\alpha$ values require the rate of decrease to be similarly large to prevent a target switch. These large decreases in uncertainty happen for at most a few time steps which leads to frequent switching, low clipped integral, and high fuel usage. When $\alpha$ is small, the opposite occurs and agents stay on targets for extended periods of time which saves fuel at the cost of lowered clipped integral performance. Finally, the performance of the task allocation algorithm can be evaluated using the results of the preceding experiments. For this comparison, the CBBA algorithm is set with the following parameters: planning depth of one, a discount factor of 0.1, and an $\alpha$ of 0.1. 

\begin{figure}[H]
    \centering
    \includegraphics[width=0.75\textwidth]{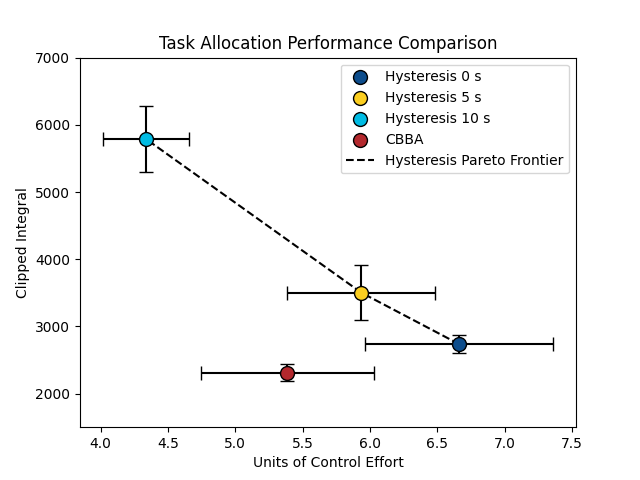}
    \caption{Scatter plot of mean performance for different algorithm configurations over a range of random initial conditions. For each configuration, 100 simulations were run, and the 95\% confidence interval is shown for each point with error bars.}
\end{figure}

As expected, increasing the hysteresis value decreases fuel expenditure, but increases the clipped integral. This is because larger hysteresis values mean less frequent target switches which saves fuel. However, this comes at the cost of the agents waiting a long time to get updates about some objects. This tradeoff between fuel and clipped integral forms a Pareto frontier which is marked in Figure 1 with a dashed line. For an algorithm to improve upon the hysteresis approach, its average must be below the current frontier. This is exactly the new approach based on the CBBA algorithm achieves meaning that this approach represents a better overall algorithm according to these performance metrics. Not only that, but this algorithm also has better clipped integral performance than the best hysteresis configuration.

Based on these results, it is clear that one of the main advantages of this approach is that it weights observation targets by their angular distance from the current sensor pointing direction. Figure 2 demonstrated that adding a small discount factor has the potential to decrease fuel use at no cost to clipped integral performance. The other modifications to the algorithm are more difficult to isolate, however, the new switching function and its scoring metric likely account for the rest of the improvements seen in Figure 4.
\section{Conclusions}

This paper presents a novel task allocation algorithm for local catalog maintenance in multi-satellite systems. It utilizes the CBBA algorithm to coordinate decentralized task allocation between agents, makes use of a new scoring function to determine the observation quality of a given target, and uses the rate of decrease of this score to inform target switching. Relevant tunable parameters of the algorithm were varied to demonstrate their effect on overall performance. Reasonable parameters were found for the discount factor and $\alpha$ for the case of two agents and eight bodies. Note that these are likely to change depending on the number of agents and bodies, the capabilities of the sensors, agent attitude control capabilities, and the initial state of the system. Additionally, the new algorithm was compared against the current approaches to demonstrate that it outperforms the current Pareto frontier.

There are several possible directions for future improvements to this work. One option is to score and observe collections of targets that all fall within the FOV of the sensor. This would allow agents to intentionally observe several targets simultaneously which could improve performance on both metrics over the course of the simulation. Another option is abandoning the fixed uncertainty requirement so that the target switching is purely based on the tradeoff between information and fuel. This could limit some of the issues with planning depth since there would not be a need for a target blacklist. In terms of simulation fidelity, using an algorithm that can handle asynchronous communication such as ACBBA could make these results more applicable to real decentralized systems. Also of interest for practical applications would be a better understanding of the convergence properties of CBBA or ACBBA for directed and time-varying graphs. Finally, an ML approach could be used to perform task allocation and or control. This could allow for complex behaviors that would otherwise be difficult to explicitly define.






\bibliographystyle{AAS_publication}
\newpage
\bibliography{fullpaper}

@article{hays2024,
author = {Hays, Christopher W. and Henderson, Troy and Miller, Kristina and Phillips, Sean and Soderlund, Alexander},
title = {Angles-Only Cooperative Local Catalog Maintenance of Close-Proximity Satellite Systems},
journal = {Journal of Guidance, Control, and Dynamics},
volume = {47},
number = {12},
pages = {2573-2586},
year = {2024},
doi = {10.2514/1.G008280},

URL = { 
    
        https://doi.org/10.2514/1.G008280
    
    

},
eprint = { 
    
        https://doi.org/10.2514/1.G008280
    
    

}
}

@ARTICLE{choi2009,
  author={Choi, Han-Lim and Brunet, Luc and How, Jonathan P.},
  journal={IEEE Transactions on Robotics}, 
  title={Consensus-Based Decentralized Auctions for Robust Task Allocation}, 
  year={2009},
  volume={25},
  number={4},
  pages={912-926},
  doi={10.1109/TRO.2009.2022423}}

@inproceedings{phillips_case_2021,
	address = {VIRTUAL EVENT},
	title = {A {Case} {Study} on {Auction}-{Based} {Task} {Allocation} {Algorithms} in {Multi}-{Satellite} {Systems}},
	isbn = {978-1-62410-609-5},
	url = {https://arc.aiaa.org/doi/10.2514/6.2021-0185},
	doi = {10.2514/6.2021-0185},
	language = {en},
	urldate = {2025-06-05},
	booktitle = {{AIAA} {Scitech} 2021 {Forum}},
	publisher = {American Institute of Aeronautics and Astronautics},
	author = {Phillips, Sean and Parra, Fernando},
	month = jan,
	year = {2021},
}

@inproceedings{johnson_asynchronous_2011,
	address = {St. Louis, Missouri},
	title = {Asynchronous {Decentralized} {Task} {Allocation} for {Dynamic} {Environments}},
	isbn = {978-1-60086-944-0},
	url = {https://arc.aiaa.org/doi/10.2514/6.2011-1441},
	doi = {10.2514/6.2011-1441},
	language = {en},
	urldate = {2025-06-05},
	booktitle = {Infotech@{Aerospace} 2011},
	publisher = {American Institute of Aeronautics and Astronautics},
	author = {Johnson, Luke and Ponda, Sameera and Choi, Han-Lim and How, Jonathan},
	month = mar,
	year = {2011},
}

@inproceedings{buckman_partial_2019,
	title = {Partial {Replanning} for {Decentralized} {Dynamic} {Task} {Allocation}},
	url = {http://arxiv.org/abs/1806.04836},
	doi = {10.2514/6.2019-0915},
	urldate = {2025-06-05},
	booktitle = {{AIAA} {Scitech} 2019 {Forum}},
	author = {Buckman, Noam and Choi, Han-Lim and How, Jonathan P.},
	month = jan,
	year = {2019},
	note = {arXiv:1806.04836 [cs]},
}

@inproceedings{aguilar_jaramillo_decentralized_2025,
	address = {Orlando, FL},
	title = {Decentralized {Consensus}-{Based} {Algorithms} for {Satellite} {Observation} {Reactive} {Planning} {With} {Complex} {Dependencies}},
	isbn = {978-1-62410-723-8},
	url = {https://arc.aiaa.org/doi/10.2514/6.2025-1148},
	doi = {10.2514/6.2025-1148},
	language = {en},
	urldate = {2025-08-11},
	booktitle = {{AIAA} {SCITECH} 2025 {Forum}},
	publisher = {American Institute of Aeronautics and Astronautics},
	author = {Aguilar Jaramillo, Alan and Gorr, Ben J. and Gao, Huilin and Mehta, Ankur and Sun, Yizhou and Ravindra, Vinay and David, Cedric and Allen, George and Selva, Daniel},
	month = jan,
	year = {2025},
}

@article{CWH_Equations,
author = {CLOHESSY, W. H. and WILTSHIRE, R. S.},
title = {Terminal Guidance System for Satellite Rendezvous},
journal = {Journal of the Aerospace Sciences},
volume = {27},
number = {9},
pages = {653-658},
year = {1960},
doi = {10.2514/8.8704},
URL = {
        https://doi.org/10.2514/8.8704
},
eprint = { 
    
        https://doi.org/10.2514/8.8704
}
}

@inbook{bertuccelli_UAV,
    author = {Luca Bertuccelli and Han-Lim Choi and Peter Cho and Jonathan How},
    title = {Real-Time Multi-UAV Task Assignment in Dynamic and Uncertain Environments},
    booktitle = {AIAA Guidance, Navigation, and Control Conference},
    chapter = {},
    pages = {},
    doi = {10.2514/6.2009-5776},
    URL = {https://arc.aiaa.org/doi/abs/10.2514/6.2009-5776},
    eprint = {https://arc.aiaa.org/doi/pdf/10.2514/6.2009-5776}
}

@inbook{lee_agile_observation,
    author = {Minjoon Lee and Sung Jun Kim and Ho-Yeon Kim and Han-Lim Choi},
    title = {Consensus-based Task Scheduling Algorithm for Agile Earth Observation Satellites with Different Authorities},
    booktitle = {ASCEND 2021},
    chapter = {},
    pages = {},
    doi = {10.2514/6.2021-4122},
    URL = {https://arc.aiaa.org/doi/abs/10.2514/6.2021-4122},
    eprint = {https://arc.aiaa.org/doi/pdf/10.2514/6.2021-4122}
}

@Article{zhang_UAV,
AUTHOR = {Zhang, Yaozhong and Feng, Wencheng and Shi, Guoqing and Jiang, Frank and Chowdhury, Morshed and Ling, Sai Ho},
TITLE = {UAV Swarm Mission Planning in Dynamic Environment Using Consensus-Based Bundle Algorithm},
JOURNAL = {Sensors},
VOLUME = {20},
YEAR = {2020},
NUMBER = {8},
ARTICLE-NUMBER = {2307},
URL = {https://www.mdpi.com/1424-8220/20/8/2307},
PubMedID = {32316556},
ISSN = {1424-8220},
DOI = {10.3390/s20082307}
}

@INPROCEEDINGS{UKF_formation_flying,
  author={Nemati, Mohammad Hossein and Kankashvar, MohammadRasoul and Bolandi, Hossein},
  booktitle={2022 30th International Conference on Electrical Engineering (ICEE)}, 
  title={Unscented Kalman Filter adaptive noise covariance selection for satellite formation flying with Q\_leaming}, 
  year={2022},
  volume={},
  number={},
  pages={362-367},
  doi={10.1109/ICEE55646.2022.9827301}}

@article{q_learning_tumbling_inspection,
author = {Aurand, Joshua and Cutlip, Steven and Lei, Henry and Lang, Kendra and Phillips, Sean},
title = {Deep Q-Learning for Decentralized Multi-Agent Inspection of a Tumbling Target},
journal = {Journal of Spacecraft and Rockets},
volume = {61},
number = {2},
pages = {341-354},
year = {2024},
doi = {10.2514/1.A35749},

URL = { 
    
        https://doi.org/10.2514/1.A35749
},
eprint = { 
    
        https://doi.org/10.2514/1.A35749
}
}

@inbook{space_construction,
author = {A. Scott Howe and Silvano Colombano},
title = {The Challenge of Space Infrastructure Construction},
booktitle = {AIAA SPACE 2010 Conference \&amp; Exposition},
chapter = {},
pages = {},
doi = {10.2514/6.2010-8619},
URL = {https://arc.aiaa.org/doi/abs/10.2514/6.2010-8619},
eprint = {https://arc.aiaa.org/doi/pdf/10.2514/6.2010-8619}
}

@article{NOACK201735,
title = {Decentralized data fusion with inverse covariance intersection},
journal = {Automatica},
volume = {79},
pages = {35-41},
year = {2017},
issn = {0005-1098},
doi = {https://doi.org/10.1016/j.automatica.2017.01.019},
url = {https://www.sciencedirect.com/science/article/pii/S0005109817300298},
author = {Benjamin Noack and Joris Sijs and Marc Reinhardt and Uwe D. Hanebeck}
}

\end{document}